# PICKY EATERS MAKE FOR BETTER RATERS

Sasha Stoikov
Cornell Financial Engineering Manhattan
2 W Loop Rd, New York, NY 10044
sfs33@cornell.edu

Stefano Borzillo
EHL Hospitality Business School
HES-SO, University of Applied Sciences and Arts Western Switzerland
Route de Cojonnex 18
1000 Lausanne 25, Switzerland
stefano.borzillo@ehl.ch

Steffen Raub
EHL Hospitality Business School
HES-SO, University of Applied Sciences and Arts Western Switzerland
Route de Cojonnex 18
1000 Lausanne 25, Switzerland
steffen.raub@ehl.ch

**Abstract**

It has been established in the literature that the number of ratings and the scores restaurants obtain on online rating systems (ORS) significantly impact their revenue. However, when a restaurant has a limited number of ratings, it may be challenging to predict its future performance. It may well be that ratings reveal more about the user who did the rating than about the quality of the restaurant. This motivates us to segment users into "inflating raters", who tend to give unusually high ratings, and "deflating raters", who tend to give unusually low ratings, and compare the rankings generated by these two populations. Using a public dataset provided by Yelp, we find that deflating raters are better at predicting restaurants that will achieve a top rating (4.5 and above) in the future. As such, these deflating raters may have an important role in restaurant discovery.

# 1. Introduction

Online Rating Systems (ORSs) have transformed how consumers decide what items to purchase, movies to watch, hotels to stay in, wine to drink, and restaurants to eat at. In theory, by democratizing the act of criticism, ORSs have challenged gatekeepers of the past and opened a path of discovery to a myriad of products and services. In practice, however, the last 20 years have seen the rise of superstars that have dominated these ORSs and made it harder for new entrants to gain traction. Restaurants are no exception to this phenomenon.

Can restaurant ratings be trusted? Answering this seemingly simple question requires considering the stakeholders involved in the rating process, i.e., restaurants, users, and ORSs. Each has incentives to inflate their rating. Firstly, *restaurants* interested in increasing their reputation may be tempted to inflate their scores through fraud or questionable methods (Luca and Zervas, 2015; Wang *et al.,* 2022, Zhang *et al.,* 2022; Zhu *et al*., 2019). Secondly, *users* fulfill their tendency to rate more things they like than things they dislike (Day and Kumar, 2023; Park and Nicolau, 2015; Stamolampros and Korfiatis, 2018). Finally, *ORSs* may benefit from the fact that high ratings lead to more consumption (Aziz et al., 2020). The actions of all of these stakeholders lead to inflated ratings which may or may not be deserved. Inflated rates thus make it hard to distinguish between truly exceptional restaurants and those that benefit from inflated ratings.

In order to distinguish the truly exceptional restaurants, there is a need for more critical opinions. Like Michelin critics, users who are hard to please – which we refer to as "deflating raters" - may be better suited to uncover restaurant quality than the average rater. It is possible that within an ORS, a subset of "deflating" raters may deflate the restaurants whose ratings are less deserved and boost the reputations of lesser-known gems. We are interested in identifying these deflating raters and demonstrating their value in predicting future restaurant scores.

In the domain of restaurant ratings, which are fundamentally linked to the taste of the users, the role of user heterogeneity is likely to be important. Not only do users disagree on what makes a good or a bad restaurant experience, but they differ on the statistics of their ratings. In this paper, the statistic we focus on is the average rating per user, which indicates how "picky" that user is. This paper will define "inflating raters" as the users who give unusually high ratings and "deflating raters" those who give unusually low ratings. Intuitively, a 4-star rating does not mean the same thing to a deflating rater with an average rating of 3 stars as to an inflating rater with an average rating of 5 stars. Therefore, every rating is driven by both the restaurant's quality and the user's average rating. In this context, our first research objective is thus to *understand whether restaurant ratings say more about the user who did the rating or about the quality of the restaurant*.

After highlighting the importance of the users and segmenting them into inflating and deflating raters, we will be interested in making a value judgment on these two populations. Thus, our second research objective is to *determine whether deflating raters are better than inflating raters.* In order to answer this question, we will define what makes a good user in terms of their ability to predict future scores. Since restaurants with a high number of ratings and high scores have been shown to perform better in terms of revenue, this is a good proxy for the future performance of restaurants.

## 2. Literature Review

### 2.1 Economic impact of ratings

Why are online ratings important to the economy? An important strand of the literature on restaurant and hotel-related ORSs focus on the impact of ratings on the *economic performance* of the businesses being rated. Combining ratings data from ORSs and business performance data (from public records or directly from ORSs), extant research has established a strong correlation between the two. Luca (2016), Aziz *et al.* (2020), and Sayfuddin and Chen (2021), for instance, found that higher-rated businesses are associated with higher revenues. Studies have also shown that other performance-related metrics such as net sales, guest count, and average check (Kim *et al*., 2016), reservation availability (Anderson and Magruder, 2011), profitability (Wang *et al*. 2021), transaction size (Torres *et al*. 2015) and restaurant closure (Yoo and Suh, 2022) are affected by ratings. What emerges from all these studies is that both the quantity of ratings and average score for a business positively impact the business's bottom line.

### 2.2 Review usefulness

What makes a single review useful? An important feature of most ORSs is that they give users the ability to rate the usefulness of reviews. After reading a review, a user may respond to the question, "Was this review helpful?". An extensive empirical literature using Yelp and

TripAdvisor reviews has emerged around identifying features that explain what constitutes a useful or helpful review.

The first *feature* to attract attention was the *numerical value of the rating itself.* Lee *et al.* (2011) found that negative reviews were typically deemed more useful than positive reviews. This has been confirmed and refined by subsequent multiple studies (Filieri *et al.*, 2019; Kwok and Xie, 2016; Lee *et al.,* 2017; Li *et al.,* 2019; Liang *et al.,* 2019; Yang *et al*., 2017). In a somewhat similar vein, Park and Nicolau (2015) have shown that *extreme ratings*, i.e., 1 or 5 stars, are considered more useful than moderate ratings.

A second feature authors focused on is the *content of the review* to uncover what constitutes a helpful review. Using text mining approaches to uncover the emotional aspects of the review text, Li *et al*. (2020) and Wang *et al.* (2019), for instance, found that reviews conveying *negative emotions* are more useful than those containing positive emotions. On their side, Liu and Park (2016) introduced metrics based on qualitative aspects of the review text, such as the *readability* of the text. They found such metrics to be most influential on the review's usefulness. Just as interestingly, Ma *et al.* (2018) studied the *inclusion of photos* in the review and found them helpful when combined with the review text.

Thirdly, extant studies have also widely reported that *features related to the reviewer* are important in assessing reviews' usefulness. These studies have focused on the reviewer's specific features, such as *elite status* (Hlee *et al.,* 2019; Kwok and Xie, 2016; Li *et al*., 2017; Li *et al*., 2019; Liang *et al*. 2019), *years of experience on the ORS* (Kwok and Xie, 2016), *number of cities visited* (Kwok and Xie, 2016; Liang et al. 2019), and *number of followers* (Hlee et al. 2019; Lee, Kwon and Back, 2021). In general, the better the reputation of a reviewer, the more useful other users deem the review.

Two recurrent themes (related to the reviewer) emerge from what has been described above: the *importance of negative opinions* and the *reputation* of the reviewer. This suggests that the most useful reviews on Yelp and TripAdvisor come from critical, experienced reviewers who are sparing with their positive reviews. These users, much like Michelin critics, may be most qualified to identify true gems in the restaurant landscape.

*2.3 Bias in the rating process*

There has been a significant interest in the biases that can affect the reliability of ratings given by online consumers (e.g., Gao et al., 2015; Hu et al., 2017; Han and Anderson, 2020; Jiang and Guo, 2015; Li and Hitt, 2008; Wang and Anderson, 2023). Among these biases, the *self-selection bias* is particularly prominent, as participating in ORS is a fundamentally voluntary activity. Self-selection bias occurs when a specific category of consumers, is more inclined to write reviews. This phenomenon may lead to a distortion in the representation of average opinions, with published reviews failing to capture the full range of experiences of all users. Hu et al. (2017) have broken down the self-selection bias into two distinct biases: the *underreporting bias* and the *acquisition bias*. These two biases play a key role in explaining the J-shaped distribution of online product reviews. Underreporting bias contributes to the bimodality of the distribution while acquisition bias is responsible for the asymmetry and tendency to inflate ratings.

*Underreporting bias* occurs when consumers with moderate or average experiences are less likely to write reviews, leading to a lack of representativeness of "ordinary" experiences in overall reviews. The underreporting bias in online reviews, which reflects the under-representation of moderate experiences, is closely related to the *propensity to rate bias* identified by Han and Anderson (2020). The latter points out that extreme, particularly negative reviews, are more frequently published, especially by users less accustomed to the review process. Together, these biases explain the tendency of online reviews to lean towards more extreme ratings, accentuating the predominance of negative reviews, especially for consumers unfamiliar with review platforms. This motivates our focus on raters with some experience, i.e. more than 5 ratings, on the Yelp platform.

*Acquisition bias* occurs when consumers with positive expectations or exceptional experiences are more inclined to acquire a product or service and, consequently, to leave reviews. This can often lead to inflated ratings, as only consumers with a positive expected net utility tend to write a review (Hu et al., 2017). However, this tendency towards inflated ratings can be ambiguous: these reviews may reflect an authentic and faithful assessment of the consumer experience or an overconfidence in their ability to choose good products or services. It is important to recognize that inflated ratings are prevalent in the context of online reviews and their potential detrimental impact has been widely documented.

*2.4     Inflated ratings: Why they are detrimental and how they can be mitigated*

It has been widely observed that ORSs have implausibly high ratings, and researchers have expressed concern that it may be difficult to distinguish between individuals or organizations that are truly exceptional and those that benefit from inflated ratings (Filippas *et al*., 2022; Garg and Johari, 2019; Horton and Golden, 2015)**.** Several empirical studies have established the prevalence of inflated ratings. On eBay, more than 90% of sellers studied between 2011 and 2014 had a rating of at least 98% positive (Nosko and Tadelis, 2015). On the online freelancing platform oDesk, average ratings rose by one star over seven years (Filippas *et al*., 2022; Horton and Golden, 2015). On taxi platforms like Uber and Lyft, inflation is particularly extreme, as anything less than 5 stars puts a driver at risk of deactivation (Athey *et al.*, 2019). On Airbnb, almost 95% of hosts have an average rating of 4.5-5 out of 5 stars (Zervas *et al*., 2021). On Amazon, ratings tend to be bimodal, with the biggest peak at 5 stars and a small peak at 1 star (Hu *et al*., 2009).

Recent research has looked at *strategies* for mitigating the occurrence of inflated ratings. A natural experiment on Bookings.com, where the rating system changed from a smiley rating to a 10-point rating, revealed a change in the distribution of ratings (Kim *et al*. 2022; Leoni and Boto-Garcia, 2023), hinting at the role of the ratings interface on mitigating the occurrence of inflated ratings. There have been some proposed solutions to making ratings less inflated. Firstly, it has been suggested to make feedback private (Horton and Golden 2015). Secondly, a randomized control trial on an online labor market showed that positive-skewed verbal scales (putting words like "the best I have ever hired") significantly reduce the occurrence of inflated ratings (Garg and Johari, 2019). Thirdly, a randomized control trial on a music discovery platform showed that introducing timers on the higher end of the rating scale can encourage user pickiness (Shahout *et al*., 2023).

Our research focuses on a post-processing approach, where we select subsets of the user population who provide less inflated ratings. In other words, we identify clusters of users who naturally submit lower average ratings. We then examine how well the ratings of these users predict restaurant quality.

## 3. Methodology

*3.1 Dataset (Yelp)*

We use data from the publicly available Yelp academic dataset and the code and results in this paper can be entirely reproduced here. We use three provided tables: the review table, the business table, and the user table.

The *review table* has 7M reviews, out of which 4.7M reviews have the "restaurant" tag. These restaurant reviews have the following structure:

1. User_id
2. Business_id
3. Rating (1-5 star)
4. Other data, like review text, date, usefulness, etc.

The *business table* has 52k businesses with the following structure:

1. Business_id
2. Rating count
3. Yelp Score (average rating rounded to the nearest half-integer)
4. Categories, which include the "restaurants" tag.
5. Other data, like location, hours, food type, etc.

The *user table* has 1.4M users with the following data structure:

1. User_id
2. Rating count
3. The average rating given by that user
4. Other data, like number of friends, user name, elite status, etc.

To our best knowledge, the phenomenon of inflated ratings has not been investigated on Yelp though the average user rating in the dataset of 3.79 suggests that ratings tend to be higher than the objective quality of the restaurants, which we expect to be closer to 3, if the average restaurant is to be calibrated to the average score on the 1-5 scale. We constructed four figures for a better understanding of the Yelp dataset. Figure 1 – Panel 1 displays the Yelp score distribution for restaurants rated between 10 and 199, and restaurants rated between 200 and 2000 times. Both distributions peak at a 4.0 Yelp score. For restaurants rated a modest amount of times, in the 10-199 range, there are a significant number of restaurants with a 3.0 or less.

However, once a restaurant has been rated more than 200 times, the scores mostly fall within the 3.5-4.5 range. Notably, the Yelp dataset has approximately 5000 restaurants with more than 200 ratings.

Figure 1 – Panel 2 displays the average rating distribution for users in the 0-4 ratings range and the 5-2000 ratings range. We note that users with very few ratings produce ratings that are very polarized, with a majority of 5-star ratings and a large peak at 1-star ratings. After about 5 ratings, the distribution starts looking like the restaurant distribution: with a peak at 4 stars. This indicates that the underreporting bias is likely to be mitigated once a rater is familiar with the platform. However, the peak at 4 stars indicates that acquisition bias may still play an important role. We define "deflating" raters to be those in the bottom 25th percentile of users who rated 5 or more restaurants. Similarly, we define "inflating" raters as those in the top 75th percentile of users who rated 5 or more restaurants.

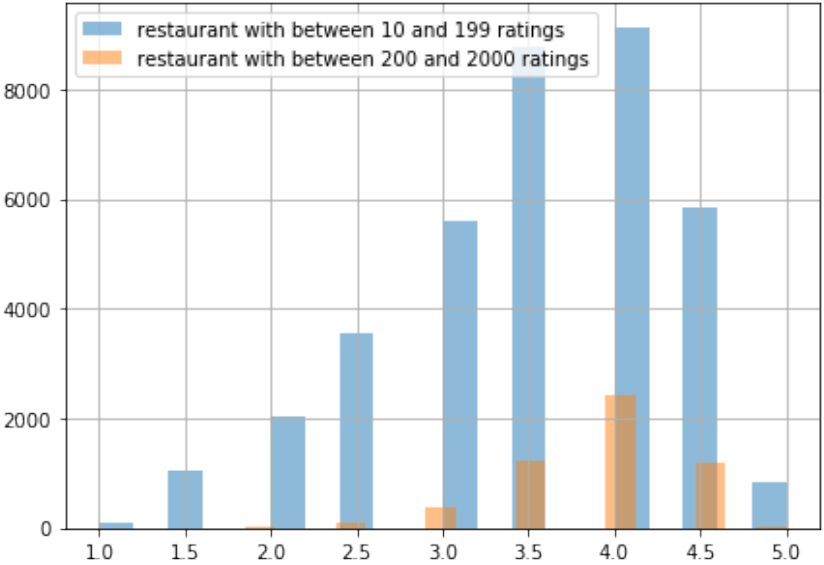

Figure 1 – Panel 1 – Yelp score distribution for restaurants with differing quantity of ratings

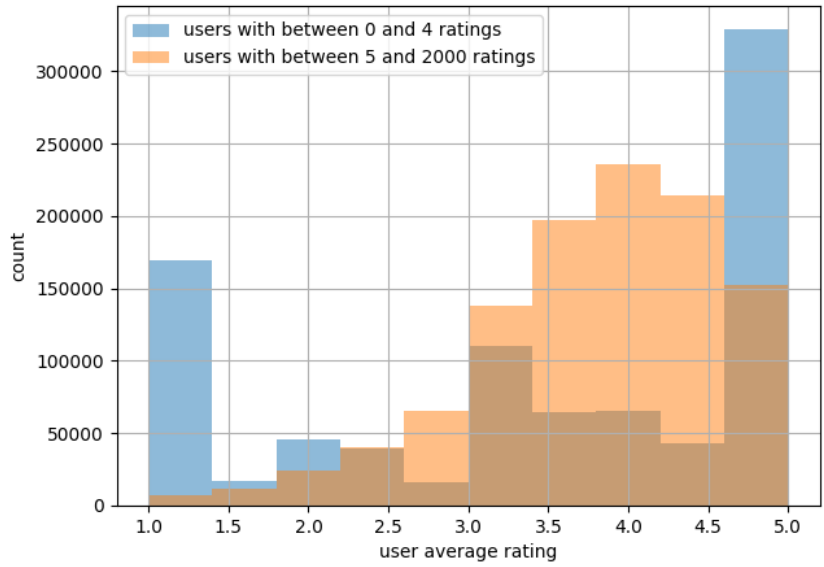

*Figure 1 – Panel 2 – Rating distribution for raters with differing quantity of ratings*

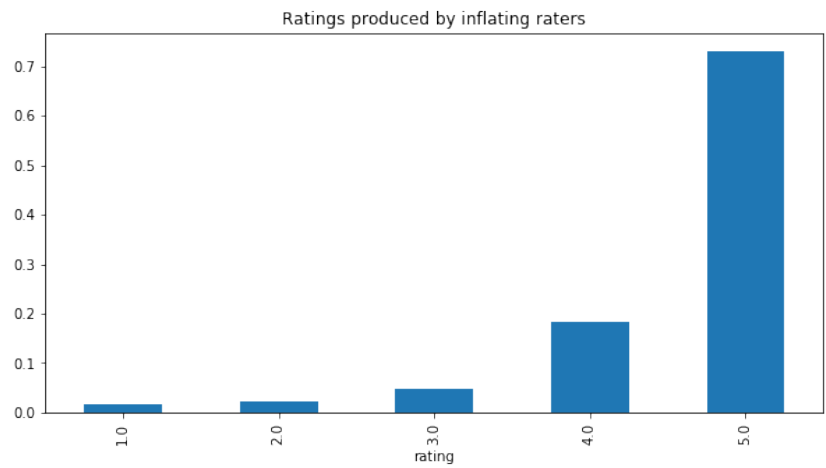

*Figure 2 – Panel 1 – Ratings from inflating raters*

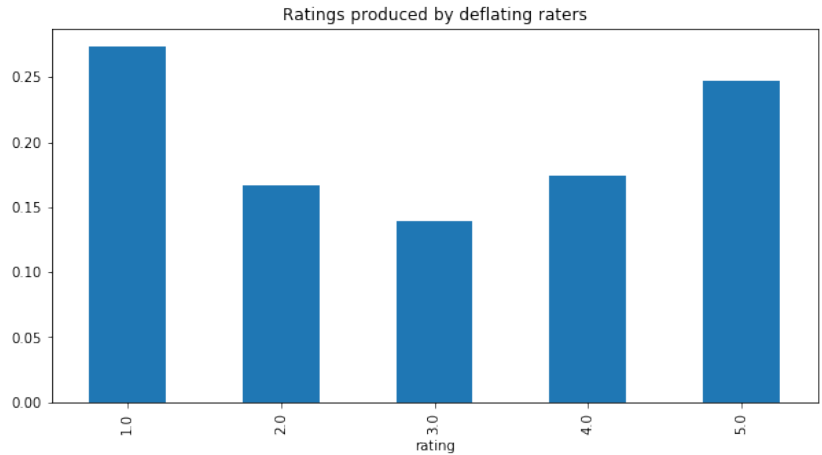

*Figure 2 – Panel 2 – Ratings from deflating raters*

Figure 2 – Panel 1 shows the distribution of ratings from inflating raters, i.e., users with an average rating greater than the 75th percentile threshold of 4.33. Figure 2 – Panel 2 shows the distribution of ratings from deflating raters, i.e., users with an average rating lower than the 25th percentile threshold of 3.38. The distributions of inflating and deflating user's ratings are very different, with the deflating users producing an almost uniform distribution of ratings, while inflating users overwhelmingly dole out 5 star ratings. Inflating users have a high mean score and low variance. Deflating users have high variance and a mean centered around 3. As we will see in our findings in section 5, these statistics have significant downstream consequences on the ratings and rankings of restaurants.

## 3.2 Regression analysis

Do ratings say more about the user who did the rating or about the quality of the restaurant? We use linear regressions, similar to much of the literature on rating usefulness (described in subsection 2.2 of our literature review) to explore this question. The main difference in our regression is that the dependent variable is the numerical value of the rating, rather than a rating usefulness metric. The independent variables are normalized average user rating and normalized restaurant Yelp score. The relative values of the coefficients indicate whether a single rating is more a product of the user's average behavior or the restaurant's average rating.

We expect the impact of the average user rating to get averaged out and disappear as a restaurant receives more ratings. This motivates us to run a regression of average restaurant ratings

(independent variable) against average user rating for that restaurant (dependent variable), for restaurants rated in the 200-2000 times range, to see if the relationship between average restaurant ratings and average user ratings persists.

This method differs from most of the literature on rating usefulness in that it involves averaging over multiple ratings on a per-restaurant basis. The two variables may be correlated, but the direction of causality is debatable. It may be that restaurants with higher ratings attract a higher proportion of inflating raters simply because these raters are better at finding quality restaurants. It is also plausible that restaurants with higher ratings are simply better are motivating inflating raters to rate and find ways to motivate deflating raters to abstain. Indeed, one of the limitations of ratings datasets is that the true quality of a restaurant is not observable, making it challenging to assess the relative qualities of various types of raters.

In what follows, we will assume that for restaurants rated more than 200 times, the Yelp score is a proxy for quality, though the role of marketing in incentivizing inflating raters needs to be acknowledged.

### 3.3 Yelp score prediction

Our second research question concerns the relative value of raters. Are deflating raters better at predicting Yelp scores than inflating raters? Ideally, we want to know which group is more accurate in predicting the "true" quality ranking of the restaurants. Since this true restaurant quality is not observable, and arguably, does not even exist, we will need to use a proxy. In practice, we use the Yelp score (a simple average of all ratings for a restaurant, rounded to the nearest 0.5) for restaurants that have been rated more than 200 times and have been rated by at least 50 inflating raters and 50 deflating raters. The overwhelming majority of restaurants that fulfill these conditions have a rating of 3.5, 4.0 and 4.5 and we define our universe to be the 3146 restaurants that fall in one of these three categories. We find 768 restaurants with a rating of 3.5 (22% of restaurants), 1697 restaurants with a rating of 4 (54% of restaurants), and 681 restaurants with a rating of 4.5 (24% of restaurants). Thus, a random ranking of restaurants would have an accuracy of 22% in predicting restaurants with a 3.5 rating, 54% in predicting restaurants with a 4.0 rating and 24% in predicting restaurants with a 4.5 rating.

The method we use generally falls under the family of bootstrapping methods. The bootstrap method is a statistical technique for estimating quantities about a population by averaging estimates from multiple small data samples. Samples are constructed by drawing observations from a large data sample with replacement. In our case, we are sampling from inflating and deflating raters to produce a ranking of our restaurant universe. This ranking is then used to classify restaurants into the low (3.5), middle (4.0) and high (4.5) categories and compared to the "true" ranking, given by the average of all ratings, i.e. the Yelp score. Alternatively, we could exclude inflating and deflating raters from this average to compute the "true" ranking, but since the results were very similar, we focus on the Yelp score as a target to compute accuracy.

The entire process for bootstrapping, classifying restaurants and the subsequent computation of accuracy scores can be summarized as follows:

1. For each restaurant in our universe and each user category (either "inflating" or "deflating" raters):
    a) Sample 20 users with replacement.
    b) Compute the average restaurant score.
2. Sort the restaurant by average user score.
3. Classify restaurants in the low (3.5), middle (4.0), and high (4.5) score categories, using percentile thresholds from the random baseline. In other words, the top 24% or restaurants are classified as 4.5, the next 54% are classified as 4.0 and the bottom 22% are classified as 3.5.
4. Compute the accuracy of this classification with respect to the Yelp score based on all ratings. In other words, the accuracy score represents the proportion of restaurants from the bootstrapping procedure that are correctly slotted into the score category to which they belong when all user ratings are considered.

By repeating the above procedure multiple times, we can compute an average accuracy as well as standard error bars. These error bars allow us to determine with confidence, whether deflating raters significantly outperform inflating raters.

## 4. Results

*4.1 Regression analysis*

With the purpose of explaining user-restaurant ratings at the single rating level we estimated the following equation:

$$y_{ij} = a + b_1 X_{1i}^{norm} + b_2 X_{2j}^{norm} + e_{ij}$$

where:

$y_{ij}$ = rating by user i of restaurant j

$X_{1i} = \frac{1}{n_i}\sum_j y_{ij}$ = user i's average rating where $n_i$ is the number of ratings by user i

$X_{2j} = \frac{1}{n_j}\sum_i y_{ij}$ = restaurant j's average rating where $n_j$ is the number of ratings for restaurant j

The normalized user and restaurant ratings are given by

$$X_{1i}^{norm} = \frac{X_{1i} - \mu_u}{\sigma_u} = \text{normalized user rating}$$

and

$$X_{2j}^{norm} = \frac{X_{2j} - \mu_r}{\sigma_r} = \text{normalized restaurant rating}$$

where $\mu_u$ and $\sigma_u$ are the mean and standard deviations of all users' average ratings and $\mu_r$ and $\sigma_r$ are the mean and standard deviations of all restaurants' average ratings.

Results of the regression are displayed in Table 1.

|  | Coefficient | Stand. error | t-value | p-value |
|---|---|---|---|---|
| Intercept | 3.7938 | 0.000 | 7598.20 | 0.000 |
| Average normalized user rating | 0.6617 | 0.001 | 1281.539 | 0.000 |
| Average normalized restaurant rating | 0.4214 | 0.001 | 816.131 | 0.000 |
|  | **R²** | 0.391 |  |  |
|  | **Adj. R²** | 0.391 |  |  |
|  | **F-value** | 1.520e+06 |  |  |
|  | **p-value** | 0.000 |  |  |
|  | **df residuals:** | 4724461 |  |  |
|  | **df model:** | 2 |  |  |

*Table 1 – Regression analysis – DV = average user rating of the restaurant*

The constant of this regression is 3.8, representing the average restaurant rating on Yelp. The normalized average user rating coefficient of 0.66 (p < .001) represents the impact of an increase or decrease of one standard deviation of the average user rating. This is significantly larger than the normalized restaurant rating average coefficient of 0.42 (p < .001), which represents the impact of an increase or decrease of one standard deviation of the restaurant score. The R squared of that regression is 39%.

Our next regression focuses on restaurants with significant amounts of ratings, in the 200 to 2000 ratings range. The independent variable is now the average rating for each restaurant and the dependent variable is the average rating of raters for that restaurant.

$Y_j = a + b\, X_{2j}$

$Y_j = \frac{1}{n_j} \sum_{\{users\ i\ who\ rated\ restaurant\ j\}} X_{1i}$ = average user rating, for users who rated restaurant j

Where $X_{1i}$ and $X_{2j}$ are the average user rating and the average restaurant rating. Note that the index j ranges over restaurants rated between 200 and 2000 times. Figure 3 displays a scatter plot of the relationship between the average restaurant rating and the average user rating, with the regression line.

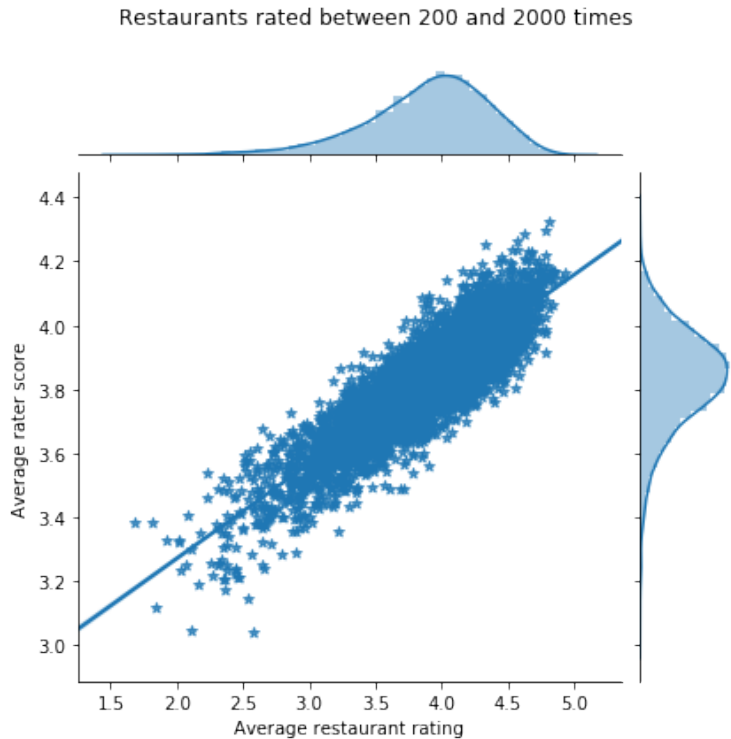

*Figure 3 – A scatter plot of the relation between average restaurant rating (not rounded) vs average rater score for that restaurant*

The slope of the regression is 0.30, indicating that for each additional star a restaurant has, the average rating of their raters goes up by 0.30 stars. The R squared of this regression is 76%, indicating that a restaurant's score is significantly influenced by the type of raters that did the rating, even for restaurants rated more than 200 times.

Are highly rated restaurants good at attracting inflating raters or are inflating raters good at finding highly rated restaurants? The average rating of inflating raters and deflating raters are 4.7 and 2.7, respectively. However, the average rating of restaurants rated by these two groups are 4.0 and 3.7, respectively. This regression to the mean indicates that the effect whereby inflating raters are better at choosing good restaurants is relatively modest. The difference between the 4.0 and 3.7 averages may be partially explained by the inflating raters' own direct influence on the average rating.

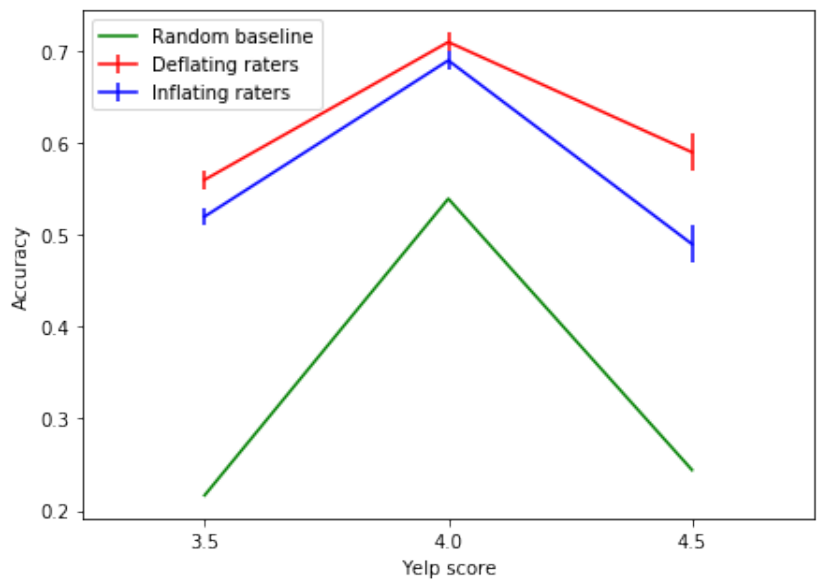

*Figure 4 – Accuracy scores for random baseline vs. deflating and inflating raters*

*4.2 Yelp score prediction*

Figure 4 displays the main result of our paper. The horizontal axis indicates the score categories (i.e. 3.5, 4.0 and 4.5) and the vertical axis represents accuracy scores. We have included a random baseline to compare to the accuracy scores for deflating and inflating raters. Sampling deflating raters to sort the restaurants in our universe of 3000 restaurants produces rankings that are better predictors of Yelp scores than inflating raters, for the top and bottom quartile of restaurants. For instance, in the bottom quartile the accuracy score for deflating raters is 56% (compared to 52% for inflating raters) and in the top quartile, the corresponding scores are 59% for deflating raters (compared to 49% for inflating raters). In the two central quartiles, the accuracy rates of both types of raters are similar and possibly not significant, as the error bars indicate. The numerical score coming from deflating raters is lower than that of inflating raters, which is why restaurants do not tend to like these users, much like the fear a restaurant may feel toward a hard-nosed Michelin critic. However, these results show that these users are more informative when it comes to producing reliable rankings.

## 5. Discussion

### 5.1 Implications for theory

Our study primarily enriches existing research on strategies for mitigating the negative consequences of inflated ratings. Extant research has focused mostly on possible changes to the rating system, including changes in the ratings interface (Garg and Johari, 2019, Kim *et al*. 2022; Leoni and Boto-Garcia, 2023, Shahout *et al*., 2023). In contrast to these upstream approaches, our findings suggest that the impact of inflated ratings can be mitigated via a "post processing" approach. By filtering raters and selecting a subset of deflating raters, the predictive accuracy of ratings, in particular at the top and bottom quartiles of the distribution can be improved. This suggests that the data provided by deflating raters may be of higher value than that of inflating raters.

One possible limitation of our result is that taking the ratings of restaurants rated more than 200 times as a proxy for quality may be questioned. The strong correlation between these restaurant's scores and the average behavior of their raters (figure 3) could just mean that restaurants with a 4.5 score are better at attracting the kind of raters who are generous. However, since there is significant trust in systems like Yelp, we can reasonably expect that restaurants with more than 200 ratings are closer to an idealized objective notion of quality.

Our results also touch indirectly on the economic impact of ratings. If ratings are a true reflection of restaurant quality, then we expect that in a fair environment, higher rated restaurants will do better economically. Our study therefore highlights reasons why the link between ratings and economics is less than perfect. First, we find that the average rating of a restaurant can often be explained in terms of the generosity or pickiness of the raters. Though this may be particularly harmful to restaurants with few ratings, we find that this phenomenon persists, even for restaurants rated more than 200 times. Second, we find that ratings inflation can weaken the link between ratings and the economic performance of a restaurant.

*5.2    Managerial implications*

Our findings have a number of interesting implications for managerial practice. In the following section we will discuss those from the perspective of ORS platforms and from the viewpoint of restaurant managers.

*5.2.1   Implications for ORS platforms*

Our findings suggest that not all restaurant raters are created equal. In particular, we find that "deflating" raters, i.e. those who display a tendency to be more critical in their average ratings of all the restaurants they have assessed, are better at predicting whether a particular restaurant belongs to the top or the bottom end of the distribution. In other words, as compared to the raters that are of the "inflating" kind, "deflating" raters seem to be more discerning experts that are better at identifying the "true gems" and the "duds", two categories of restaurants that users of ORS platforms particularly care about.

For ORS platforms this would provide the opportunity to compute a differentiated restaurant score based exclusively on the ratings of the more critical "deflating" raters. Such a differentiated scoring system would bear some similarity with the "audience score" vs. the "critics score" on Rotten Tomatoes, a popular review-aggregation website for film and television. Whereas the current Yelp score would reflect the more inflating "audience" score, with ratings that are generally biased upwards and condensed at the upper end of the rating scale, an additional "Yelp critics" score based on the ratings of "deflating" raters would have more in common with the "critics" score which reflects the opinion of more discerning expert judges.

From a guest perspective, such a new score would arguably provide much richer information. The current problem with ORS scores is a general tendency towards grade inflation, leading more and more users to rely exclusively on the content of the comments, rather than the overall numerical rating. A new score based on critical raters would have greater diagnostic value in and by itself. In particular, it would help guests single out promising restaurants that are at the top of the distribution, and avoid those that are at the lower end.

*5.2.2 Implications from the restaurant perspective*

For restaurateurs, a "Yelp critics" scores could also be beneficial. While one could argue that restaurants care above all for high average ratings, they also suffer from the fact that grade inflation reduces an individual restaurant's ability to stand out from the crowd. A high score on "Yelp critics" would therefore be a more potent marketing argument compared to the current scoring system.

In addition, if restaurants would receive information not only about their average score but also about the average scores of the raters that assessed them, this would allow for a prediction of the direction in which their score is likely to evolve in the future. As Figure 3 suggests, average restaurant ratings and average rater scores are highly correlated. Therefore, if a restaurant knows that the average scores of its raters are higher than its average score, regression to the mean would suggest that in the future the average rating of this restaurant will go down, as over time a higher proportion of critical raters will weigh in. Conversely, if the average scores of its raters are lower than its average score, an improvement of the average restaurant score is to be expected over time.

The latest argument is particularly important for restaurants with limited numbers of ratings. When such restaurants, that may have been opened recently, have received a low rating and the raters are particularly picky, their scores are likely to pick up substantially as more raters are added.